\documentclass[english, 12pt]{article}
\usepackage[T1]{fontenc}
\usepackage{babel}
\usepackage{ifsym}
\usepackage{geometry}
\usepackage{amsmath, amsthm, amssymb}
\usepackage{graphicx}
\usepackage{booktabs}
\usepackage{rotating}
\usepackage{setspace}
\usepackage{authblk}  
\usepackage{epigraph}
\usepackage{enumitem}   

\usepackage[round]{natbib}
\usepackage[colorlinks,
linkcolor=red,
anchorcolor=blue,
citecolor=blue
]{hyperref}

\newtheorem*{theorem*}{Theorem}
\newtheorem*{corollary}{Corollary}
\newtheorem{lemma}{Lemma}

\newtheorem{Result}{Result}

\newcommand{\R}{\mathbb{R}}

\newcommand{\pard}[2]{\frac{\partial #1}{\partial #2}}
\makeatletter
\usepackage{babel}

\makeatother
\usepackage{footmisc}
\makeatother

\begin{document}
\title{Domestic Constraints in Crisis Bargaining}
\pagenumbering{gobble}
\author{Liqun Liu\thanks{Ph.D. student, University of Chicago Harris School. 1307 E 60th Street, Chicago, IL 60637. Email: liqunliu@uchicago.edu.}}
\maketitle
\begin{abstract}
\begin{doublespacing}
I study how political bias and audience costs impose domestic institutional constraints that affect states' capacity to reach peaceful agreements during crises.  With a mechanism design approach, I show that the existence of peaceful agreements hinges crucially on whether the resource being divided can appease two sides of the highest type (i.e. the maximum war capacity). The derivation has two major implications.  On the one hand, if war must be averted, then political leaders are not incentivized by audience costs to communicate private information; they will pool on the strategy that induces the maximum bargaining gains.
On the other hand, political bias matters for the scope of peace because it alters a state's expected war payoff.

		\end{doublespacing}
			\vspace*{1cm}
			
			\noindent \textbf{Word Count:} 3704
			\vspace*{1cm}
			\end{abstract}
	\clearpage
\doublespacing
\pagenumbering{arabic}
\section{Introduction}
A near-consensus in the international relations literature is that information asymmetry could give rise to war when states engage in crisis bargaining \citep{B88,Fearon95}; and, mediation is  valuable because it reduces the chances that states miscalculate in the face of this information asymmetry. An oft-cited example is Robert Gates's successful mediation of the 1990 Indo-Pakistani crisis in Kashmir. According to \cite{H98}, underlying the success are Gates's persuasive message to both sides that the war would be detrimental, and his promise to monitor and share information about states' compliance using American spy satellites.

But states rarely bargain in vacuum. More often than not, when leaders bargain on behalf of domestic constituencies, they have misaligned preferences for war and peace. The formal international relations literature has identified two sorts of domestic constraints that create such an agency issue. One constraint is called {\it political bias}  \citep{Bdm99, JM07}: during conflicts, leaders might gain/lose disproportionally relative to their fellow citizens. The other constraint is called {\it audience costs}: when the peril of war has been precluded,  political leaders incur a penalty by domestic constituencies because they back down from crisis \citep{Fearon94, Sc, AL12}. We ask: when these domestic constraints exist, how do they interact with information asymmetry and thereby determine the scope of peace in crisis bargaining?

{\bf My approach.} I cast the question in the context of meditated crisis bargaining. I augment the canonical crisis bargaining games  (see also \cite{Banks90} and \cite{FR11}) with political bias and audience costs, and examine their equilibrium properties with the game-free approach. Following \cite{AR17}, I model political bias as a cost parameter that differentiates the war payoffs for leaders and citizens within a state; this approach captures the idea that people do not equally benefit or suffer from a conflict. I impose no restrictions on the functional form of audience costs.  Given an arbitrary audience cost, we say a  {\it peaceful mechanism} exists whenever there is a settlement preferable to the expected war outcomes for two sides. A crisis bargaining situation is {\it peace-implausible} whenever peaceful mechanisms do not exist for {\it any} audience cost; it is {\it peace-plausible} otherwise.

This approach helps us identify the limits of peaceful resolution subject to domestic constraints. When peace is implausible, conflicts must happen with strictly positive probabilities regardless of how audience costs are chosen. But if peace is plausible, there might be ways to resolve the crisis peacefully.   To illustrate why the approach establishes a ``possibility'' result, let us assume that a crisis is due to be mediated in the real world. In the most benign situation, a Gates-like powerful mediator might step in. She then proposes and commits to enforcing the settlement specified by the peaceful mechanism as an ultimatum to two sides; war ensues conditional on no deal. After some domestic processes, states might want to agree on the settlement, because they cannot ask for more by threatening to use force.

{\bf Main results.} I show that the plausibility of peace hinges on a ``weighted budget constraint'': when the total size of resource being divided exceeds the sum of two highest-type states' expected war payoffs, peacefully resolving the crisis is possible.  

From the characterization, one immediately recognizes that  political bias matters for the scope of peace. The rationale is straightforward: war is states' alternative means  to resolve crises other than signing a peaceful agreement. Since political bias describes how much each side may obtain by exercising the war option, it affects the plausibility of peaceful resolution.

Perhaps surprisingly, audience costs do not show up in the characterization. Indeed, the result comes from the delicate interaction between peace plausibility and audience costs: from any peaceful mechanism, political leaders must receive payoffs net of audience costs constant across private types. To see why, let's
take for granted a powerful mediator who is determined to preclude war possibilities. Per earlier discussions, she is able to implement any peaceful settlement whenever there is one. Now we ask: might the leader of a given state 
gain/lose by saying different things to the mediator? No -- if the mediator's goal is to preclude war possibilities, then she cannot penalize leaders by letting them fight for whatever they say. This means that leaders of all types would pool by communicating in a way that induces the highest net payoff from peace, which is lower-bounded by what a strongest type would expect from conflicts (where political bias matters). 

The characterization of peace plausibility does not mean that audience costs are unimportant; instead, knowing whether peace is plausible allows us to further explore the right kind of audience costs across different contexts in crisis bargaining situations. In mediation, characterizing peace plausibility helps us identify institutional constraints that would fundamentally limit states' capacities to resolve the crisis peacefully. If peace is indeed plausible, then a powerful mediator who wants to end crises might achieve her goal by encouraging states to design audience costs appropriately. When averting war is not a must, audience costs are again incredibly useful institutional arrangements to discipline political leaders' behaviors.  From this perspective, this paper is complementary to \cite{AR17}. They analyze the optimal domestic constraint design during crises from the standpoint of a state, while I study the possibility of peaceful resolution when domestic constraints are real concerns.

Methodologically, this paper relates to the mechanism design approach to war. Along this line, \cite{Banks90} first recognizes the necessity of an ``institution-free'' analysis for IR models. \cite{FR09, FR11} study the possibility of peaceful resolution. \cite{HMS} point out that if a mediator communicates to two sides cleverly, sometimes mediation could achieve the same resolution outcome (not necessarily peaceful) with arbitration. \cite{FK21} show that the ultimatum game is a robust protocol in studying crisis bargaining situations. \cite{KS21} reexamine received wisdom by adding low-level costly actions to the canonical models. This paper studies the limits of peaceful resolution subject to domestic constraints; it also generalizes a set of results in the existing literature, such as the weak monotonicity results between resolve and war propensity \citep{Banks90} and the characterization of peaceful mechanisms \citep{FR09, FR11}.

\section{Model}\label{M}
There are two states disputing over a unit size of resource. Each state has a population normalized to 1 with share $\gamma_i$ being leaders and share $1-\gamma_i$ being citizens. Leaders  bargain in the shadow domestic constraints. At the outset, there is a mediator whose objective is to preclude war.\\

\noindent{\bf Mediated crisis bargaining.} Leaders engage in a crisis bargaining game, whose final outcome is either a peaceful settlement or a war. The game has the standard structure of information uncertainty about resolve\footnote{See \cite{R17} for a review.}; resolve affects states' war payoff but not their preferences for settlements. Let the type space be $ \Theta=\Theta_1\times\Theta_2$ where $\Theta_i$ is an interval $[\underline{\theta}_i, \bar{\theta}_i]$. A typical element within $\Theta$ is $\theta=(\theta_1, \theta_2)$, in which $\theta_1$ and $\theta_2$ are independently distributed. Each state knows its own type $\theta_i$, but treats state-$j's$ type as a random variable with cumulative distribution function $F_j(\theta_j)$ and strictly positive density.

The crisis bargaining outcome is a tuple $(\pi, x)$ specifying war probability $\pi\in [0, 1]$ and peaceful division $x=(x_1, x_2) \in \Delta^2=\{(x_1, x_2):x_1+x_2\leq 1, x_1, x_2\in \R_+\}$. War is a costly lottery with each side endowed with winning probability $p=(p_1, p_2)$ which sums to one. Leaders also suffer from war costs $c=(c_1, c_2)$ adjusted by their political biases. The winner takes all the resources. 

$\theta$ determines state-$i$ leader's war payoff $w_i(\theta)$, which is strictly increasing and differentiable in its argument $\theta_i$. According to \cite{FR11}, the resolve $\theta_i$ may be the (one-sided) fighting costs $w_i(\theta)=w_i(-c_i)$ where $c_i\sim F_i(c_i)$ and $w_i$ is increasing in $-c_i$; it may also be the (two-sided) relative strength in fight $w_i(p_i(\theta))=w_i(p_i(\theta_i, \theta_j))$ where $p_i$ is increasing in $\theta_i$ but decreasing in $\theta_j$, and $w_i$ is increasing in $p_i$.
 
Denote leaders' strategy set abstractly as $S=S_1\times S_2$. $S$ itself is enormously large, but it contains a small subset $\Theta=\Theta_1\times\Theta_2$ which coincides with leaders' type space. The class of game form known as the {\it direct mechanism} restricts each leader's strategy to reporting some type $\tilde{\theta}_i\in \Theta_i$; it assigns war probability $\pi(\tilde{\theta})$ and peaceful settlement $x(\tilde{\theta})=(x_1(\tilde{\theta}), x_2(\tilde{\theta}))$ for a typical report profile $\tilde{\theta}\in \Theta$. Formally, we represent the direct mechanism as a ``menu'' $\Sigma: \tilde{\theta}\rightarrow (\pi(\tilde{\theta}), x(\tilde{\theta}))$. It is incentive compatible (IC) if leaders of all types are willing to report truthfully (i.e. $\tilde{\theta}_i=\theta_i$). By the powerful revelation principle \citep{M79,FR11}, if  our goal is to study the peace possibility, then restricting attention to the class of incentive compatible direct mechanisms $\Sigma$ entails no loss. 
 
At this point, the common game-theoretic approach would be positing an exact game form so that we may analyze its equilibrium outcome. The following game form $\Gamma$ is an example: 
\begin{enumerate}
    \item The mediator offers and commits to a ``menu'' $\Sigma$ to the leaders of two states-in-conflict. 
    \item Each leader sends a report $\tilde{\theta}_i$ to the mediator.
    \item The outcome $(\pi(\tilde{\theta}), x(\tilde{\theta}))$ realizes according to the menu.
\end{enumerate}
If $\Sigma$ is incentive compatible, then we should expect $(\pi(\theta), x(\theta))$ to emerge as an equilibrium outcome. 

Per the revelation principle, there is a deep connection between the game form $\Gamma$ and the equilibrium outcome of {\it any} crisis bargaining game. Precisely, the equilibrium outcome of an arbitrary crisis bargaining game takes the form $(\pi(\theta), x(\theta))$;
a powerful mediator may implement this outcome according to $\Gamma$ by offering an incentive compatible menu $\Sigma: \tilde{\theta}\rightarrow (\pi(\tilde{\theta}), x(\tilde{\theta}))$ to two sides. This alludes the existence of an incentive compatible and peaceful menu $\Sigma$ as the limits of peaceful resolution: 
if no incentive compatible menu $\Sigma$ averts war surely, then
{\it no} crisis bargaining outcome can be peaceful.   \\

\noindent{\bf Modeling domestic constraints.} No matter whether the crisis ends up in war or peace, leaders might receive a payoff different from citizens'. I consider two sorts of domestic constraints, audience costs and political bias. \\

\noindent{\it Audience costs.} Formal literature \citep{Fearon94, Sc, AL12, AR17} considers audience costs as leaders' electoral costs if they initiate crises but end up backing down. Along this line, we may view audience costs as the part of payoff  that are privately ``valued'' by political leaders; they give rise to agency issues conditional on peaceful settlements. 

The most standard and intuitive approach to capture this idea is as follows: the leader pays an electoral cost ($a_i<0$) or obtains an electoral benefit  ($a_i>0$) depending on the deal $x_i$ she/he has struck from the bargaining table ($x_i$ includes the status quo).  Since $x_i$ is commonly valued by the leader and citizens, we may write their respective payoffs from the peaceful settlement as $v_i=x_i-a_i$ and $x_i$ (or $v_i(\theta)$ and $x_i(\theta)$, if we want to emphasize the dependence on the private information $\theta$). 

There is a mathematically equivalent view of audience costs that is easier to work with for this paper's purpose: we consider audience costs as mappings from each peaceful settlement $x_i$ to the payoff differential $a_i$ between leaders and citizens.  To see the equivalence, fix the audience costs and a settlement $x_i$ from the bargaining table. Conditional on this settlement, a typical citizen simply gets $x_i$; the leader obtains a net payoff $v_i\neq x_i$ because audience costs are a real thing. From an outsider's perspective, she may recover the audience cost $a_i$ whenever knowing $v_i$ and $x_i$ (by computing the difference). This view accommodates all plausible functional forms of $v_i$ (and therefore $a_i$), so it entails no loss of generality. Thus, I operationalize this view of audience costs and consider the pair of payoffs $v_i(\theta)$ and $x_i(\theta)$ as model primitives. \\
 
\noindent{\it Political bias.} Political bias is the war payoff discrepancy between leaders and citizens  \citep{JM07, Bdm99, AR17}. The easiest way to capture this idea is introducing a cost parameter $\lambda_i$ (as in \cite{AR17}): if a typical state-$i$ citizen pays a war cost $c_i$, then the leader pays $\lambda_i c_i$ where $\lambda_i>0$. The cost parameter $\lambda_i$ may also represent other agency issues during wartime. For example,  political elites might avoid serving military, or suffer financially in a different way than grassroots due to war. The war payoff in state $i$ would be $w_i=p_i-\lambda_i c_i$ for its leader and $w_i^c=p_i-c_i$ for each citizen. \\

\noindent{\bf Participation constraints.} Given the anarchic structure of the international system, states may back out from any signed peaceful agreement; they are willing to stay in the agreement only if doing so  brings a higher expected payoff than war. The ``voluntary participation constraints'' reflect the concern of \citet{Waltz01}, who suggests that ``sovereign states with no system of law enforceable among them, with each state judging its grievances and ambitions according to the dictates of its own reason or desire''. In the IR context, the suitable participation constraint is the ``interim individual rationality'' \citep{Banks90,FR11}. It says that stakeholders must find the assigned peaceful settlement weakly preferable to the expected war payoff.  

I consider the participation constraints for the leader and citizens respectively. Notably, considering citizens' participation constraints reflects the view that each of the domestic groups has a say in signing a bilateral peaceful settlement. Removing this constraint amounts to saying that citizens do not matter in the domestic redistribution of gains or losses out of the crisis bargaining. Were this true, the agency issue created by audience costs and political bias is gone -- after all, only the leaders' preferences matter. Of course, there are situations in which leaders have absolute authority or no autonomy in crisis negotiations. My setup accommodates these issues by setting $\gamma$ close to $1$ and $0$ respectively, interpreting $\gamma$ as leaders' weights in the domestic political processes.  

Now denote stakeholders' the expected (interim) war/peace payoff as 
\begin{align*}
(\text{War})\qquad  &W_i(\theta_i)= \int_{\underline{\theta}_j}^{\bar{\theta}_j}w_i(\theta_i,\theta_j)dF_j(\theta_j),\quad W^c_i(\theta_i)= \int_{\underline{\theta}_j}^{\bar{\theta}_j}w^c_i(\theta_i,\theta_j)dF_j(\theta_j),\\
(\text{Peace})\qquad &V_i(\theta_i)= \int_{\underline{\theta}_j}^{\bar{\theta}_j}v_i(\theta_i,\theta_j)dF_j(\theta_j), \quad X_i(\theta_i)= \int_{\underline{\theta}_j}^{\bar{\theta}_j}x_i(\theta_i,\theta_j)dF_j(\theta_j), 
\end{align*} 
Following \cite{Banks90}, I define leaders and citizens' interim participation constraints as follows: for each $\theta_i\in [\underline{\theta}_i, \bar{\theta}_i]$
\begin{align}
    (\text{Leader}) \qquad  &W_i(\theta_i)\leq V_i(\theta_i)\label{PL}\\
    (\text{Citizens}) \qquad  &W^c_i(\theta_i)\leq X_i(\theta_i)\label{PC}
\end{align}

The participation constraint is more than an artifact of the game-free analysis.  We may concretely think of its role in the standard crisis bargaining game by augmenting $\Gamma$ with a stage of domestic political processes. Let a mediator commit to offering a peaceful menu $\Sigma$ ($\pi\equiv 0$) to leaders at the bargaining table. If both leaders accept the menu, they send a type report $\hat{\theta}$ to the mediator and bring back the plan of settlement $x_i(\hat{\theta})$ for ratification.  If citizens in both states ratify the settlement, then peace realizes. War ensues whenever either the leader or citizens of any state rejects the menu.\\

\noindent{\bf Payoff structure.} Each player is risk-neutral. Let $u_i, u^c_i$ stand for the {\it ex post utility} of the leader and a typical citizen in state $i$.  Since any bargaining outcome is completely pinned down by the war probability and peace payoffs, these utilities can be represented as the average payoffs of war $(w_i, w_i^c)$ and peace $(v_i, x_i)$ weighted by the war probability $\pi$. To emphasize their dependence on the type realization $\theta\in\Theta$, we write
\begin{align*}
u_i(\theta)&=\pi (\theta)w_i(\theta)+(1-\pi(\theta))v_i(\theta)\\
u^c_i(\theta)&=\pi(\theta) w^c_i(\theta)+(1-\pi(\theta))x_i(\theta)
\end{align*}
Let $U_i(\tilde{\theta}_i|\theta_i)$ denote type-$\theta_i$ leader's expected (interim) utility if she reports $\tilde{\theta}_i$ in the mechanism,
\begin{align*}
U_i(\tilde{\theta}_i|\theta_i)=\int_{\underline{\theta}_j}^{\bar{\theta}_j} u_i(\tilde{\theta_i}, \theta_j)dF_j(\theta_j)=\int_{\underline{\theta}_j}^{\bar{\theta}_j} [\pi (\tilde{\theta}_i, \theta)w_i(\theta_i, \theta_j)+(1-\pi(\tilde{\theta_i}, \theta_j))v_i(\tilde{\theta}_i,\theta_j)]dF_j(\theta_j)
\end{align*}
and denote her utility from truthfully reporting as $U_i(\theta_i):=U_i({\theta}_i|\theta_i)$.\\

 \noindent{\bf Budget balancedness.} I also rule out the uninteresting possibility that the mediator may subsidize peace. As \cite{FR11} suggest: peace is always possible in a trivial sense  whenever a mediator is willing to make sufficiently large subsidies to appease both states. This sort of subsidies could be too large to be politically feasible.

\section{Analysis}\label{A}
\noindent{\bf Characterization.} By the revelation principle, we might without loss of generality focus on the existence of an incentive-compatible and peaceful menu $\Sigma: \theta\rightarrow (\pi(\theta), x(\theta))$ that satisfies the participation constraints and budget balancedness. $\Sigma$ is peaceful if $\pi(\theta)\equiv 0$ for all $\theta=(\theta_1, \theta_2)$ up to a measure-zero set. It is incentive-compatible if no leader at the bargaining table might gain by misreporting her/his private information $\theta_i$. 

We present two results about  incentive compatibility. The proof follows the standard mechanism design technique and is deferred to the Appendix.
\begin{lemma}\label{MONO}
Leaders' equilibrium war propensity $E_{\Theta_j}[\pi(\theta_i,\theta_j)]$ is increasing in $\theta_i$ if $w_j(\theta_i,\theta_j)$ can be written as $w_j(\theta_i,\theta_j)=h(\theta_i)-g(\theta_j)$ where $h,g$ are strictly increasing functions.
\end{lemma}
\begin{lemma}\label{env}
    $\Sigma$ is incentive compatible if and only if 
	\begin{align}
 U_i({{\theta}}_i)-U_i({\underline{\theta}_i})=\int_{\underline{\theta_i}}^{{\theta_i}}\int_{\underline{\theta}_j}^{\bar{\theta}_j} {\pi}(t, \theta_j)\pard{w_i(t, \theta_j)}{t}dF_j(\theta_j)dt, \label{Env}\\	
\int_{\underline{\theta}_j}^{\bar{\theta}_j}  {\pi}(t, \theta_j)\pard{w_i(t, \theta_j)}{t}dF_j(\theta_j)\; \text{is weakly increasing in} \;t.\label{Mo}
	\end{align}

\end{lemma}
 The lemmas jointly extend the monotonicity results of \cite{Banks90} to a setting in which domestic constraints exist. It encapsulates the risk-return tradeoff in crisis bargaining games: stronger leaders are more aggressive in initiating conflict (Lemma \ref{MONO}, Condition \ref{Mo}), and obtain higher equilibrium payoffs (Condition \ref{Env}). The underlying rationale comes from stronger leaders' ability to misrepresent private information: they can always mimic the behaviors of lower types on the bargaining table; as the better fighters, they cannot fare worse whenever conflicts happen. 

While audience costs do not explicitly enter leaders' equilibrium payoffs $U_i$, the {\it prima facie} interpretation that audience costs are payoff-irrelevant is wrong. Audience costs matter because they enter leaders' payoff $v_i$ in the absence of war; $v_i$ in turn impacts leaders' utility $U_i$. To correctly interpret the role of audience costs, we focus on its incentive effect in the class of peaceful mechanisms; there, ``peaceful'' requires  the war assignment function $\pi(\theta)\equiv 0$ for all $\theta$. Plugging in $\pi(\theta)\equiv 0$ to Conditions \ref{Env}-\ref{Mo}, we see that leaders receive constant payoffs labeled as $\bar{V}_i$ across types during peaceful mechanisms; they do not gain by misrepresenting private information. Exactly because from peaceful mechanisms, leaders' payoffs remain invariant to their private information for arbitrary audience costs design,

\begin{Result}\label{Ir}
	Audience costs are {\it irrelevant} for incentivizing peaceful settlements.	
\end{Result}

By Lemma \ref{env} and the peace requirement ${\pi}\equiv 0$, I rewrite the participation constraints. At the bargaining table, each leader understands that she/he expects a constant payoff $\bar{V}_i$ by settling down and $W_i(\theta_i)$ by fighting. In a peaceful mechanism, $\bar{V}_i$ is lower-bounded by the highest type's expected war payoff $W_i(\bar{\theta}_i)$; otherwise, the highest types prefer to fight. Hence
 leaders' participation constraints must satisfy the following inequality:
\begin{align*}
(\text{Leader})\qquad &W_i(\bar{\theta}_i)\leq \bar{V}_i \quad \forall \theta_i\in [\underline{\theta_i}, \bar{\theta}_i]
\end{align*}

Similarly, rewriting citizens' participation constraints  as follows:
\begin{align*}
    (\text{Citizens})\qquad &W^c_i(\theta_i)\leq X_i(\theta_i)\quad \forall \theta_i\in [\underline{\theta_i}, \bar{\theta}_i]
\end{align*}
Turn to the budget-balancedness requirement and fix the realization $\theta_i$. At this interim stage, the leader and citizens of state $i$ cannot accept a settlement  inducing a pair of payoffs worse than $\bar{V}_i$ and $X_i(\theta_i)$. In total, appeasing state $i$ costs at least
\begin{align*}
   R_i(\theta_i)=\gamma_i \bar{V}_i+(1-\gamma_i)X_i(\theta_i)
\end{align*}
for each $\theta_i$. When a mediator designs the menu $\Sigma$, she/he must prepare for the worst situation i.e. two strongest leaders meet at the bargaining table. This is the moment when the price of peace is maximal  $(R_i(\bar{\theta}_i))$. To preclude war possibility, the budget balancedness requires $\sum_i R_i(\bar{\theta}_i)\leq 1$. It together with the participation constraints gives us the peace-plausibility condition:
\begin{align}
\text{(Peace-plausibility condition)}\qquad \sum_{i} \gamma_i W_i(\bar{\theta}_i)+(1-\gamma_i)W^c_i(\bar{\theta}_i)\leq 1\label{RC}
\end{align}

\begin{Result}\label{Exi}
A peaceful mechanism exists if and only if the peace-plausibility condition holds. In this case, there exists a pair of audience costs that guarantees the existence of a peaceful settlement.
\end{Result}
The peace-plausibility condition should come at no surprise: in vein of \cite{FR11}, it technically says that peace is possible when the total size of resource being divided is large enough to satisfy two states. But only political bias enters this condition, audience costs do not. 

To understand this asymmetry, recall the channels that political bias and audience costs affect leaders' war-peace decisions. Political bias matters because it affects the expected war payoff of highest types; if they cannot be pacified, war must happen with strictly positive possibilities. Audience costs matter to the extent that leaders' peace payoff $\bar{V}_i$ is pinned down, which is lower-bounded by what a highest-type leader could obtain in war. To gain more intuitions, we may view political bias as prescribing the minimal size of pie that a state would demand to refrain from using force; audience costs describe how a state may redistribute the pie between the groups of leaders and citizens -- thanks to audience costs,  leaders obtain a fraction $\gamma_i v_i$ ($v_i=x_i\Leftrightarrow a_i=0$) while citizens obtain a fraction $(1-\gamma_i) x_i$ from any peaceful settlement $x_i$.   

The peace-plausibility condition has implications for the design of optimal domestic constraints in crisis bargaining situations. When it fails, at least the leader or citizens from a particular state would refuse to settle down. In other words, the (violation of the) peace condition makes a case in which citizens {\it cannot} design an audience cost that prevents the war possibility; it  resolves a theoretical puzzle in \cite{AR17} as to why the war possibility is independent of the domestic preference divergence (measured by audience costs)\footnote{``Interestingly, the decision about whether or not to use a strategy with positive probability of war is independent of the preference divergence between the leader and the citizen.'' \cite[14]{AR17}}.

From the peace-plausibility condition, we can also sharpen a set of robust comparative static predictions known as the ``monotonicity results'' in the classic game-free analysis \citep{Banks90, FR11}. Absent any peaceful mechanism, stronger leaders' excessive war propensity manifests itself by a strictly positive war probability. This means that the monotonicity result must be strict over some range of types.
\begin{corollary}
	If the peace-plausibility fails, then in any equilibrium of the crisis bargaining game there exist strong-type leaders obtaining strictly higher payoffs.
\end{corollary}

\section{Conclusion}\label{C}
This paper analyzes the role of domestic constraints in general crisis bargaining situations.  With the game-free approach, I derive the ``weighted budget constraint'' as necessary for plausibility of peaceful resolution. The condition implies that political bias matters for the scope of peace, because it affects how much states may take away from the bargaining table by threatening to go to war. If war  must be averted, then audience costs  never induce an equilibrium bargaining outcome that might vary with private information. In the language of contract theory, the audience costs design does not provide incentives for leaders to communicate private information within peaceful mechanisms; their major role  is domestically redistributing spoils for each peaceful settlement.

This paper provides several benchmark results for future research on crisis bargaining models. For example, I extend the monotonicity results from \cite{Banks90} to a setting in which domestic constraints are real. Since these results hold regardless of the game form and domestic constraints, the value of modeling particular bargaining protocols and domestic constraints lies in generating other empirically-relevant theoretical predictions.

\clearpage{}
	\renewcommand
	\refname{Reference}
	\bibliographystyle{chicago}
	\bibliography{new.bib}
\clearpage
\appendix
\section{Appendix}
\subsection{Proof of Lemma \ref{MONO}}
We remark that this class of $w_i$ encompasses one-sided uncertainty (about war cost) and two-sided uncertainty with contest success functions in difference forms. The proof technique is standard (e.g. Lemma 1 in \cite{Banks90}). Below is a sketch: recall that $U_i(\tilde{\theta}_i|\theta_i)=E_{\Theta_j}[\pi(\tilde{\theta}_i,\theta_j)w_i(\theta_i,\theta_j)+(1-\pi(\tilde{\theta}_i,\theta_j))v_i(\tilde{\theta}_i,\theta_j)]$. Let $\theta_2>\theta_1$. Then $U_i(\theta_1|\theta_1)\geq U_i(\theta_2|\theta_1)$ and $U_i(\theta_2|\theta_2)\geq U_i(\theta_1|\theta_2)$. Adding up, we have $E_{\Theta_j} [\pi(\theta_2, \theta_j)-\pi(\theta_1, \theta_j)][w_i(\theta_2, \theta_j)-w_i(\theta_1, \theta_j)]\geq 0$. Under the functional form assumption, it simplifies to $[h(\theta_2)-h(\theta_1)]E_{\Theta_j} [\pi(\theta_2, \theta_j)-\pi(\theta_1, \theta_j)]\geq 0$. Since $h$ is increasing, it must be that $E_{\Theta_j}[\pi(\theta_2,\theta_j)]\geq E_{\Theta_j}[\pi(\theta_1,\theta_j)]$. 
\subsection{Proof of Lemma \ref{env}}\label{ET}
To analyze incentive compatible mechanism, I apply the envelope theorem in \cite{MS02}.  In an IC mechanism, the leader must be willing to tell the truth. Take the most general case $w_i=w_i(\theta_i, \theta_j), v_i=v_i(\theta_i, \theta_j)$. Since truth-telling is optimal, 
\begin{align}
U_i({\theta}_i)=\max_{\hat{\theta}_i} U_i(\hat{\theta}_i|\theta_i)&=\max_{\hat{\theta}_i}\int_{\underline{\theta}_j}^{\bar{\theta}_j}[\pi(\hat{\theta}_i, \theta_j)w_i(\theta_i, \theta_j)+(1-\pi(\hat{\theta}_i, \theta_j))v(\hat{\theta}_i, \theta_j)]dF_j(\theta_j)\label{TT}
\end{align}
Since $w_i(\theta_i,\theta_j)$ is differentiable, apply envelope theorem to Equation \eqref{TT} 
\begin{align*}
\frac{dU_i({\theta}_i)}{d\theta_i}=\int_{\underline{\theta}_j}^{\bar{\theta}_j} {\pi}(\theta_i, \theta_j)\pard{w_i(\theta_i, \theta_j)}{\theta_i}dF_j(\theta_j)
\end{align*}
By the fundamental theorem of calculus,
\begin{align}
U_i({{\theta}}_i)-U_i({\underline{\theta_i}})=\int_{\underline{\theta_i}}^{{\theta_i}}\int_{\underline{\theta}_j}^{\bar{\theta}_j} {\pi}(t, \theta_j)\pard{w_i(t, \theta_j)}{t}dF_j(\theta_j)dt\label{En}
\end{align}
Equation \eqref{En} is a necessary condition for incentive compatibility. From standard Bayesian mechanism design argument (see \cite{TB15} for example), we know that it is also sufficient for incentive compatibility if $\bar{\pi}(t):=\int_{\underline{\theta}_j}^{\bar{\theta}_j}{\pi}(t, \theta_j)\pard{w_i(t, \theta_j)}{t}dF_j(\theta_j)$ is nondecreasing in $t$.

\subsection{Proof of Result \ref{Exi}}
\begin{proof}
 If the condition fails, then for any settlement we can identify a nontrivial subset of $\Theta$ that contains at least one unsatisfied stakeholder. To see it, when two strongest types $\bar{\theta}_1, \bar{\theta}_2$ meet, they cannot simultaneously agree on any peaceful settlement simply because doing so would require a budget larger than permitted,
\begin{align*}
     \sum_i R_i(\bar{\theta}_i)&= \sum_i \gamma_i \bar{V}_i+(1-\gamma_i)X_i(\bar{\theta}_i)\\
     &\geq \sum_{i} \gamma_i W_i(\bar{\theta}_i)+(1-\gamma_i)W^c_i(\bar{\theta}_i)>1
\end{align*}
Since $W_i, W_i^c$ are continuous in $\theta_i$, the set of pairs $(\theta_1, \theta_2)$ with $\theta_i$ close enough to $\bar{\theta}_i$  will for the same reason refuse to sign any peaceful settlement. Because this set of ``near-strongest duos'' is non-degenerate, war will happen with strict positive probabilities. 

If instead this condition holds, a division $(X_1, X_2)$ satisfying $X_1\geq W_1(\bar{\theta}_1), X_2\geq W_2(\bar{\theta}_2)$ would pacify leaders and citizens alike when, for example, the audience cost is zero.
\end{proof}

\end{document}